\documentclass[pra,superscriptaddress,twocolumn]{revtex4-2}

\usepackage{amsmath}
\usepackage{physics}
\usepackage{graphicx}
\usepackage[colorlinks=true,linkcolor=blue,citecolor=red,urlcolor=magenta]{hyperref}
\usepackage{charter}

\begin{document}

\begin{abstract}
    The investigation of many-body interactions holds significant importance in both quantum foundations and information. Hamiltonians coupling multiple particles at once, beyond others, can lead to a faster entanglement generation, multiqubit gate implementation and improved error correction. As an increasing number of quantum platforms enable the realization of such physical settings, it becomes interesting to study the verification of many-body interaction resources. In this work, we explore the possibility of higher-order couplings detection through the quantum Fisher information. For a family of normalised symmetric $k$-body Ising-like Hamiltonians, we derive bounds on the quantum Fisher information in product states. Due to its ordering with respect to the order of interaction, we demonstrate the possibility of detecting many-body couplings for a given Hamiltonian from the discussed family by observing violations of an appropriate bound.
    As a possible extension to these observations, we further analyse an example concerning the three-body interaction detection in the XY model.
\end{abstract}

\title{Exploring Many-body Interactions Through Quantum Fisher Information}

\author{Pawe{\l} Cie\'sli\'nski}
\email{pawel.cieslinski@phdstud.ug.edu.pl}
\affiliation{Institute of Theoretical Physics and Astrophysics, University of Gdańsk, 80-308 Gda\'nsk, Poland}

\author{Pawe{\l} Kurzy\'nski}

\affiliation{Institute of Spintronics and Quantum Information, Faculty of Physics, Adam Mickiewicz University, 61-614 Pozna\'n, Poland}

\author{Tomasz Sowi\'nski}

\affiliation{Institute of Physics, Polish Academy of Sciences, Aleja Lotnikow 32/46, 02-668 Warsaw, Poland}

\author{Waldemar K{\l}obus}

\affiliation{Institute of Theoretical Physics and Astrophysics, University of Gdańsk, 80-308 Gda\'nsk, Poland}

\author{Wies{\l}aw Laskowski}

\affiliation{Institute of Theoretical Physics and Astrophysics, University of Gdańsk, 80-308 Gda\'nsk, Poland}

\maketitle

\section{Introduction}
Among many fascinating phenomena in physics, in recent years studying the nature of interactions has become not only a subject of fundamental research but also a part of the current pursuit towards modern quantum technologies. Most of the current controllable quantum systems rely solely on the two-body interactions between particles~\cite{NielsenChuang, Gross_2017,Kjaergaard_2020}. Nevertheless, many-body interactions are often discussed in the context of effective models in low-energy physics~\cite{Tseng_1999,  Peng_2009, Zhang_2022}. These include studies of spin systems~\cite{Pachos_2004, Motrunich_2005, Bermudex_2009, Muller_2011, Andrade_2022}, extended Hubbard models describing ultra-cold atoms or molecules in optical lattices \cite{2007BuchlerNatPhys,2008SchmidtPRL,2010WillNature,2011SilvaPRA,2012SafaviPRL,2012SowinskiPRA,2013MahmudPRA,2014DaleyPRA,2014SowińskiPhysScripta,2015PaulPRA,2016SinghPRA,2016HincapiePRA,2020HarshmanAnnPhys}, quantum chemistry~\cite{Aspuru_Guzik_2005, Malley_2016, Hempel_2018, Babbush_2018}
as well as nuclear and particle physics~\cite{Hauke_2013, Banuls_2020, Ciavarella_2021, Farrell_2023}. Further applications of higher-order interactions can be found in entanglement generation~\cite{Shi_2009, Peng_2010, Facchi_2011, Cieslinski_2023}, error correction~\cite{Kitaev_2003, Paetznick_2013, Yoder_2016} and others~\cite{Vedral_1996, Wang_2001, Figgatt_2017, Marvian_2022}. Thus, a search for many-body interactions plays a significant role for both, quantum foundations and future quantum technologies~\cite{Goto_2004, Monz_2009, Levine_2019, Khazali_2020, Kim_2022, Katz_2023}. With a rising demand for the implementation of many-body interaction Hamiltonians, it becomes interesting to study their verification~\cite{Marvian_2022}. A universal method solving this task could probe new physical effects and give insights on how to engineer the desired Hamiltonians. Furthermore, it would be useful for Hamiltonian learning protocols as they often require prior knowledge of the maximal degree of interaction graph (see, e.g. \cite{Anshu_2021}). On the other hand, it would answer the question of whether the Hamiltonian is even worth learning if we are interested in its many-body interaction properties.
In this work, we demonstrate that the existence of genuine many-body interactions can be verified through the quantum Fisher information (QFI), thus possibly paving the way for a new area of non-local interactions research. 

\begin{figure}
    \centering
    \includegraphics[width=0.48\textwidth]{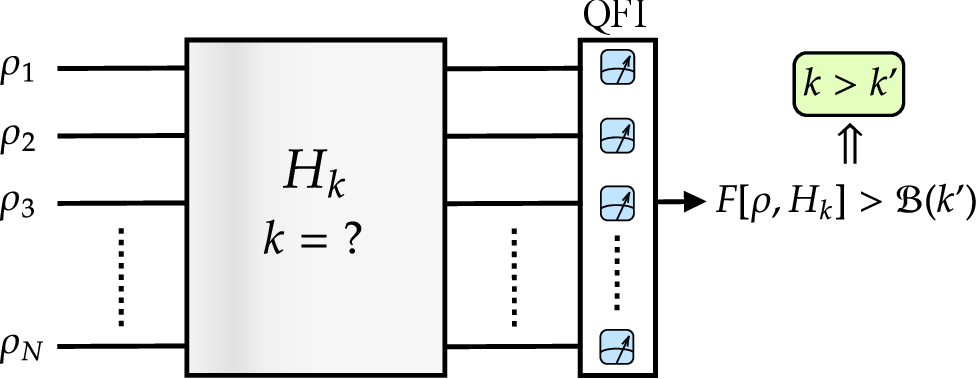}
    \caption{\textit{Motivation}. In this work, we explore the order of interactions, i.e., maximal number of simultaneously coupled qubits in Hamiltonian, through the use of quantum Fisher information. For symmetric Ising-like Hamiltonians $H_k$ ~(\ref{eq:hamiltonian_family}), we show that quantum Fisher information calculated in a product state is bounded with a bound dependent on $k$. This allows one to detect if $H_k$ manifests interactions beyond the $k'$-body case, where $k> k'$. The basic idea is the following. Starting from an arbitrary separable state we let the particles interact via $H_k$ and then we measure its quantum Fisher information. Based on our findings we check if the result is greater than the maximally allowed value for the chosen $k'$. If the answer is positive we can claim that $H_k$ contains terms of at least $(k'+1)$th order.  }
    \label{fig:motivation}
\end{figure}

QFI has been studied in various contexts~\cite{Zanardi_2008, Hyllus_2012, Toth_2012, Li_2013, Song_2013, Demkowicz_2015, Tan_2021, Alenezi_2022, Baak_2022, Baak_2023, Abiuso_2023}, including quantum phase transitions~\cite{Invernizzi_2008, Wang_2014, Song_2017, Yin_2019, Lambert_2020, Yu_2022} and most notably quantum metrology~\cite{Toth_2014, Braun_2018}.
For a given Hamiltonian it allows one to find states that guarantee measurement precision beyond the classical limit~\cite{Toth_2014, Braun_2018}. 
An experimental measurement, or estimation, of QFI can be performed via several techniques. The proposed theoretical protocols take advantage of the dynamical susceptibility~\cite{Hauke_2016}, projections onto the initial state (Loschmidt echo protocol)~\cite{Macri_2016}, overlap detection~\cite{Zhang_2017}, randomised measurements~\cite{Rath_2021, Yu_2021, Vitale_2023} and adiabatic perturbation theory~\cite{Zhang_2023}.
A direct QFI (or its lower bound) measurement was performed in, e.g. \cite{Mathew_2020,Yu_2021,Yu_2022b}.

In the standard metrological scenario, Hamiltonians under consideration are strictly local. However, many-body interacting systems were also examined in terms of improved scaling~\cite{Braun_2018}. Here, to make our presentation as simple as possible, we study the family of $k$-local permutationally invariant Ising-like Hamiltonians of $N$ particles. For them, we illustrate the main premise behind this paper, which is the ordering of maximal QFI in product states for increasing interaction order. Based on this observation we derive bounds on the Hamiltonians with at most two-body interaction terms showing the possibility of witnessing the presence of $k$-body interactions with product states, see Fig~\ref{fig:motivation}.
As a possible extension, we also discuss an example of a similar study for the XY model.

\section{Motivating example}
\label{sec:motivating_example}

At first, we will start with a simple example that motivates our work. Consider a system of three qubits on a triangle that interact in the $\sigma_z$ direction. Fixing the order of interaction at $2$, the two-body interaction Hamiltonian is given as
\begin{equation*}
    H_{2}=\frac{1}{2} \left( \sigma_z \otimes \sigma_z \otimes \openone + \openone \otimes \sigma_z \otimes \sigma_z + \sigma_z \otimes \openone  \otimes  \sigma_z \right),
\end{equation*}
where, in order to keep the correspondence with a standard metrological notation, the prefactor was chosen such that the maximal eigenvalue does not exceed $N/2=3/2$. On the other hand, focussing on the three-body couplings only, we get
\begin{equation*}
    H_{3}=\frac{3}{2} \left( \sigma_z \otimes \sigma_z \otimes \sigma_z \right).
\end{equation*}
Now, let us consider the maximisation of quantum mechanical variance $(\Delta H)^2=\langle H^2 \rangle - \langle H \rangle^2$ over the set of pure product states. For $H_{2}$ the optimal product state, i.e., the state that maximises the variance, is given as $|\psi_{prod}\rangle = \left( \sqrt{p}| 0 \rangle + \sqrt{1-p}|1\rangle \right) ^{\otimes 3}$, where $|0 \rangle$ and $|1 \rangle $ are eigenstates of $\sigma_z$ and $p=(3+\sqrt{3})/6$. See Appendix~\ref{app1} for direct calculations. The exact value of the maximal variance is $(\Delta H_{2})^2_{max}=1$. It is straightforward to show that for trivial Hamiltonians with one-body terms the variance is given as $(\Delta H_{1})^2_{max}=N/4$. Consequently, for three qubits we get $(\Delta H_{1})^2_{max}=3/4$, which is smaller than in the two-body case. Moving on to the higher-order interaction Hamiltonian the optimal state is \textit{no longer} a tensor product of identical local states $|\psi \rangle ^{\otimes 3}$. It might seem that the form of the state that maximises the variance for the two-body interaction is trivially obtained from the symmetry, but it is not. In fact, the only condition arising from it is that $\langle \psi_1 \psi_2 \psi_3 | H | \psi_1 \psi_2 \psi_3  \rangle = \langle \psi_2 \psi_1 \psi_3 | H | \psi_2 \psi_1 \psi_3  \rangle$, and similarly for all other permutations of indices. This can be seen clearly in the three-body interaction case. Examining the spectrum of $H_{3}$, we can construct a product state which is a superposition of eigenstates associated with the maximal and minimal eigenvalues, see Appendix~\ref{app1}. Namely, we have $|\psi_{prod} \rangle = |00\rangle \otimes 1/\sqrt{2}\,(|0\rangle + |1 \rangle)$ and the corresponding $(\Delta H_{3})^2_{max}=9/4$. Limiting ourselves to the product states being a tensor product of identical one-qubit states we would arrive at the value accesible for $k=2$. From these results, it is apparent that
\begin{equation*}
    \mathrm{SL}=(\Delta H_{1})^2_{max}  <  (\Delta H_{2})^2_{max} 
\end{equation*}
\begin{equation*}
     < (\Delta H_{3})^2_{max} \leftarrow \mathrm{HL},
\end{equation*}
where we denoted the values of variances coinciding with the standard limit and the Heisenberg limit for one-body Hamiltonians and entangled states as SL and HL, respectively. Here, it is also worth noting that the ordering of maximal variance is not due to the possibility of obtaining higher eigenvalues with changing interaction type or the number of terms in the Hamiltonian since they are all normalised in the same manner. This simple example shows that variance of a Hamiltonian calculated in a product state may be exploited as a tool for detecting many-body interaction terms.

\section{Physical model and preliminaries}
\label{sec:model}

In the following, to make our considerations more general, instead of the variance of a Hamiltonian we will focus on the quantum Fisher information.  For any hermitian operator $A$ and state $\rho$ it is defined as
\begin{equation}
    F[\rho, A]=2\sum_{k,l}\frac{(\lambda_k-\lambda_l)^2}{(\lambda_k+\lambda_l)}|\langle k | A | l \rangle |^2,
\end{equation}
with $|k \rangle$ and $\lambda_k$ being eigenvectors and eigenvalues of a density matrix $\rho$ respectively and the sum is evaluated only when the denominator is different from zero. Most commonly, it is used in quantum metrology as it allows one to find a suitable state that one can use to boost phase-measurement sensitivity. It was shown that for local Hamiltonians QFI in product states is bounded by $N$~\cite{ Pezze_2009}. Introducing entangled states moves this bound to $N^2$, leading to the so-called Heisenberg limit~\cite{Giovannetti_2006,Pezze_2009}. 
Moreover, for any quantum state $\rho$ there exist a pure state $|\psi\rangle$ for which \cite{Toth_2014}
\begin{equation}
    F[\rho, A] \leq 4 (\Delta A)_{|\psi \rangle }^2=F[|\psi \rangle, A].
    \label{eq:fisher_variance}
\end{equation}
Hence, any maximisation of QFI can be performed among the pure states only. This justifies the choice of variance and calculations presented in the previous section.

Our task is to construct a model and derive a bound on the quantum Fisher information in a \textit{product state} that is an increasing function of an interaction order. 
Now, let us precisely define the notion of higher-order interactions. A given Hamiltonian 
\begin{equation}
    H_k=\sum_j h_j^{(k)},
\end{equation}
is considered to be $k$-local if each of $h_j^{(k)}$ act nontrivially on at most $k$ particles. Hamiltonians which act exactly on $k$ or at most on $K$  particles will be denoted as $H_{k}$ and $\mathcal{H}^{(K)}$ respectively. By this definition, we will use the $k$-locality and $k$th order of interactions interchangeably. Measurements enhanced with Hamiltonians that are non-linear in operators have been studied in terms of sensing and scaling in the past~\cite{Braun_2018}.
It included, as an example, products of photon creation and annihilation operators~\cite{Luis_2004, Beltran_2005, Luis_2007, Woolley_2008}, angular momenta and its components~\cite{Boixo_2008, Choi_2008, Napolitano_2010, Napolitano_2011}, many-body models~\cite{Roy_2008, Ma_2009, Pezze_2016, Li_2019, Baak_2022, Baak_2023}, and more generally powers of sum of local operators~\cite{Boixo_2008b} as well as the general $k$-body~\cite{Chu_2023}, and symmetric $k$-body interaction Hamiltonians~\cite{Boixo_2008b, Czajkowski_2019} with no comparison between entangled and separable states in the latter cases.
However, in this work, we study a specific scenario, which resembles the standard metrological approach and will be used in a far different context. For such means, we will consider only the symmetric Ising-like Hamiltonians. 
First, we define the auxiliary Hamiltonians that contain only the $k$-body interaction terms as 
\begin{equation}
    H_k=\mathcal{N} \sum_{(i_1,\cdots,i_k) \in G_k} \, \sigma_z^{i_1} \sigma_z^{ i_2}  \cdots \sigma_z^{ i_N}
\end{equation}
where $\mathcal{N}$ is a normalisation constant and $G_k$ is a fully connected interaction graph for $k$-body interactions - the summation is performed over all $k$-partite subsets of particles making it permutationally invariant.
In the case of $k=1$ we retrieve the standard metrological Hamiltonians
\begin{equation}
    H_{1}=\frac{1}{2}\sum_{i=1}^N \sigma_z^i,
\end{equation}
if proper normalisation is chosen. Another example can be given for $k=2$ and $\mathcal{N}=J$, namely
\begin{equation}
    H_{2}=J \sum_{i<j}^N \sigma_z^i \sigma_z^j.
\end{equation}
Note that the above Hamiltonian represents a long-range interaction Ising model on a complete interaction graph. For more examples see Sec.~\ref{sec:motivating_example} and Sec.~\ref{sec:example}.
Now, a general symmetric Ising-like Hamiltonian containing at most $k$-body interactions can be constructed as
\begin{equation}
    \mathcal{H}^{(K)}=\mathcal{N} \sum_{k\leq K} \alpha_k H_k
    \label{eq:hamiltonian_family}
\end{equation}
where again $\mathcal{N}$ is a normalisation constant and $\alpha_k$ are real numbers. Throughout most of the paper, we will focus on $\forall_k \alpha_k=1$ and discuss its modifications in the examples.
It is worth noting that all of the results presented here hold for any Hamiltonian equivalent under local unitary operations. This follows from the property of QFI which states that $F[\rho, U^{\dagger} A U]=F[U\rho U^{\dagger}, A]$ and local unitary invariance of entanglement.
Hamiltonian from a different class will be discussed in Sec.\ref{sec:extensions}.

Before we move on to our results, we need to specify a proper normalisation for $\mathcal{H}^{(K)}$. Our motivation is to test the order of interactions present in the system. Moreover, we would like to compare our results with the metrological approach and stay consistent with its results. In order not to break the classical and Heisenberg scaling we choose to set the operator norm $||\mathcal{H}^{(K)}||=\max_{\phi} ||\mathcal{H}^{(K)} |\phi \rangle || =N/2$. Note that we want the Hamiltonian to appear as if no $k$-body interactions were present in it. Dropping this assumption we would obtain the non-linear Hamiltonians scalings $N^{\sim k}$ (see, e.g.~\cite{Luis_2004, Beltran_2005, Luis_2007, Boixo_2008, Boixo_2008b, Roy_2008, Choi_2008, Woolley_2008, Ma_2009, Napolitano_2010, Napolitano_2011, Pezze_2016, Li_2019}).
Our approach leads to an upper bound on variance and QFI which cannot exceed $N^2$ for any $k$ and quantum state $|\psi \rangle$, entangled or not. We implement this norm by setting $\mathcal{N}$ to $N/(2\max_i |E_i|)$, where $E_i$ is an eigenvalue of $\mathcal{H}^{(K)}$. As this normalisation is multiplication by a constant, there still is an overlap between the previous research and our results. This will be commented on in Sec.~\ref{sec:testing}.

\section{Interaction dependent bounds on QFI for product states}
\label{sec:ordering_of_QFI}

For the considered model it is possible to derive the explicit formulas for the eigenvalues based on its symmetry. For $k$-local Hamiltonian defined in (\ref{eq:hamiltonian_family}) and included normalisation we get
\begin{equation}
 \Omega^{N,K}_e=\sum_{k \leq K}\omega^{N,k}_e , \quad
\omega^{N,k}_e=\sum_{j=0}^e \frac{N}{2} \frac{\binom{e}{j} \binom{N-e}{k-j}}{\binom{N}{k}}
(-1)^{\,j}
\label{eq:eigenvalues}
\end{equation}
where $e$ is the number of excitations, i.e., number of $|1 \rangle$ elements in the $N$-qubit state. Here, $\omega_e^{N, k}$ represents the eigenvalues of a Hamiltonian with $k$-body terms only. 

As a first step, we will limit ourselves to a scenario with a fixed $k$, i.e. $H_k$. In such a case, the maximisation of variance and hence the QFI (\ref{eq:fisher_variance}) over pure product states can be performed as follows. Since variance is the function of the squared modulus of amplitudes and Hamiltonian eigenvalues we can consider only product states of the following form
\[
|\psi_{prod} \rangle = \bigotimes_i^N \left(\sqrt{p_i}|0 \rangle +\sqrt{1-p_i}|1 \rangle \right).
\]
Using this parametrisation and calculating the variance we get
\[ (\Delta H_k)^2 = \sum_{l_1,\cdots,l_N=0,1} \prod_{i=1}^N p_i^{l_i}(1-p_i)^{1-l_i}(\omega^{N,k}_{l_1+\cdots+l_N})^2
\]
\[
-\left(\sum_{l_1,\cdots,l_N=0,1} \prod_{i=1}^N p_i
^{l_i}(1-p_i)^{1-l_i}\omega^{N,k}_{l_1+\cdots+l_N} \right)^2.
\]
A necessary condition for the existence of multivariable function extremum is the disappearance of its first derivatives. Taking derivatives of $(\Delta H)^2$ over $p_i$ we arrive at the system of $N$ equations linear in $p_k$
\begin{equation*}
0=\frac{\partial (\Delta H_k)^2}{\partial p_k}= \sum_{l_1,\cdots,l_N=0,1} (-1)^{1-l_k} \prod_{i\neq k}^N p_i^{l_i}(1-p_i)^{1-l_i} 
\end{equation*}
\begin{equation*}
 (\omega^{N,k}_{l_1+\cdots+l_N})^2  - 2 \Bigg(\sum_{l_1,\cdots,l_N=0,1} \prod_{i=1}^N p_i^{l_i}(1-p_i)^{1-l_i}\omega^{N,k}_{l_1+\cdots+l_N}\Bigg)
\end{equation*}
\begin{equation}
\sum_{l_1,\cdots,l_N=0,1} (-1)^{1-l_k} \prod_{i\neq k}^N p_i^{l_i}(1-p_i)^{1-l_i}\omega^{N,k}_{l_1+\cdots+l_N},
\label{eq:optimisation}
 \end{equation}    
where the dependence on $p_k$ is only in the term in the parentheses. Since we know all $\omega$'s, this could be solved directly and the resulting set of $(p_1,\cdots,p_N)$ sufficing $0 \leq p_i \leq 1$ is the solution to our optimisation problem. We present a step-by-step solution for $N=3$ in Appendix~\ref{app1}.

\section{Testing \texorpdfstring{$k$}{TEXT}-body interactions}
\label{sec:testing}
The most interesting case for higher $k$-body interaction terms would be to find an explicit bound on QFI for $k=2$. For the fixed $k=1$ we once again refer to the fundamental result obtained for local Hamiltonians~\cite{Pezze_2009}. One of its many consequences is the classical scaling with $\max_{|\psi_{prod}\rangle}4(\Delta H_1)^2=N$. It is straightforward to see that our approach is consistent with that.  
By direct solutions of ~(\ref{eq:optimisation}) and numerical calculations, for $k=2$ we notice that up to $N=13$ the solution to the given optimisation problem is obtained by $|\psi_{prod} \rangle = (\sqrt{p} |0 \rangle + \sqrt{1-p} |1\rangle )^{\otimes N}$. For more discussion see Appendix~\ref{app1}. This result is known to hold asymptotically when the number of particles is much greater than the interaction order, here  $n \gg 2$~\cite{Boixo_2008b}. It is valid in our approach as a non-linear Hamiltonian  $(\sum_i H_{1})^2$ for which this result was derived is equal to $N \openone \cdots \openone +NH_{2}$ and thus holds in our case. 
\begin{figure}[ht]
    \centering
    \includegraphics[width=0.49\textwidth]{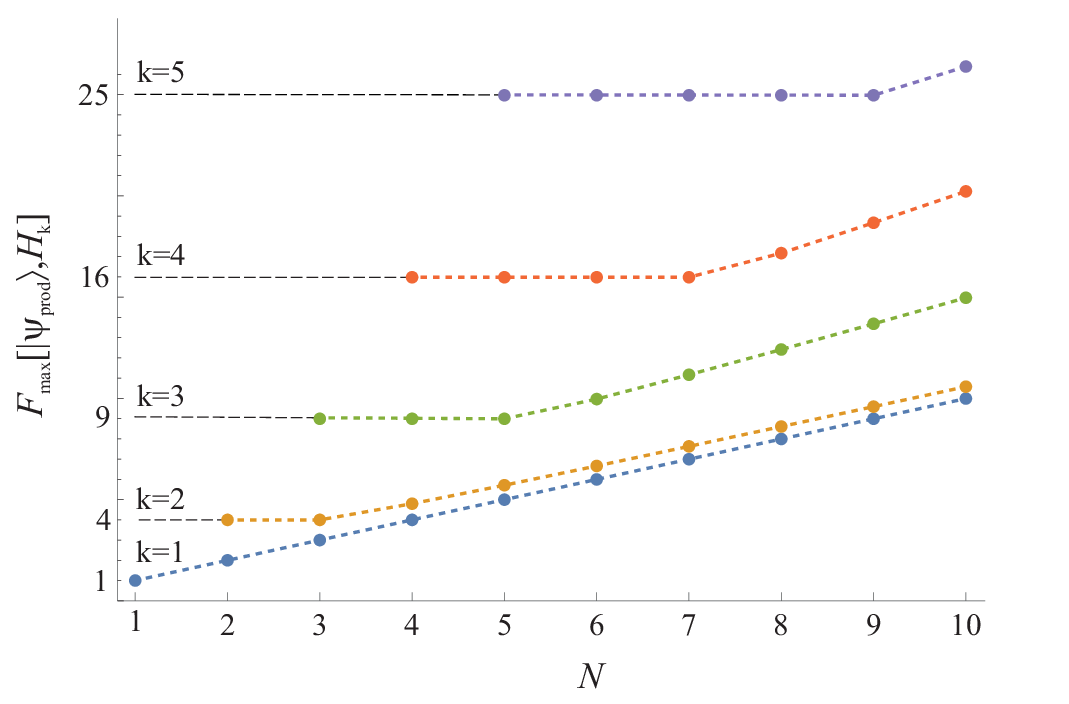}
    \caption{Maximal quantum Fisher information of $k$-local Hamiltonian $H_{k}$ in a product state as a function of the number of particles $N$. Numerical solutions of (\ref{eq:optimisation}) for different choices of $k$ are plotted in different colours, and dashed lines are guides to the eye. Values associated with $k=1$ scale linearly with $N$~\cite{Pezze_2009}, while for $k=2$ their behaviour is described by (\ref{eq:fisher_max_k2}). From the plot, one can see that there exists a clear ordering of the maximal QFI with respect to the fixed interaction order. This observation allows one to formulate criteria which would distinguish the minimal order of $k$-body interactions present in the Hamiltonian using separable states. According to our choice of normalisation when $N=k$, Heisenberg scaling of $N^2$ is achieved as expected. Note that for $N < 2k$ the trend is constant. Then, once the number of particles is equal to $2k$ the QFI grows. This change in the behaviour of QFI is necessary as it would be impossible for it to remain constant since it would eventually yield values smaller than the ones obtained with local Hamiltonians and lead to a contradiction.} 
    \label{fig:max_Fisher_fixed}
\end{figure}
Furthermore, it is consistent with the results obtained for the Lipkin–Meshkov–Glick (LMG) model and the nearest-neighbours (as well as fully connected) Ising model with interaction parameter smaller than its critical value (general parameter range)~\cite{Pezze_2016, Li_2019}. This will also be elaborated on further in the text.
For the given state, the variance reduces to 
\begin{equation*}
    (\Delta H_{2})^2=\frac{4}{N-1}\Big(-2N(2N-3)p^4+4N(2N-3)p^3
\end{equation*}
\begin{equation}
    -N(5N-7)p^2+N(N-1)p \Big).
\end{equation}
By finding zeros of its derivative we get three unique roots, from which the one that maximizes the variance is given as
\begin{equation}
    p_{max}=\frac{2 N - 3 + \sqrt{2 N^2 - 7 N + 6}}{4N - 6}
\end{equation}
The resulting maximal QFI in a product state, i.e. the solution to (\ref{eq:optimisation}) obtained with the above $p_{max}$, for $k=2$ is 
\begin{equation}
    F_{max}[|\psi_{prod}\rangle,H_{2}]=\frac{2N(N-1)}{2(N-2)+1}.
    \label{eq:fisher_max_k2}
\end{equation}
The results presented here coincide with the ones obtained for the LMG model in the limit of large coupling constant $\gamma$ and the context of statistical speed, see the Supporting Information of~\cite{Pezze_2016}. This is due to the fact that once the coupling constant is large, the single qubit terms in the LMG model can be neglected. Then, up to a constant, the LMG Hamiltonian equivalent to $H_2$ studied here.
Note that~(\ref{eq:fisher_max_k2}) is an increasing function of $N$ which bounds, from the above, the one-body scaling. Explicit results for $k~\leq~5$ and $N\leq 10$ are presented in Fig.~\ref{fig:max_Fisher_fixed}. As expected, for $k=1$ the maximal QFI scales as $N$~\cite{Pezze_2009}. For $k>1$, the results clearly show that the maximal Fisher information in a product state is ordered with respect to the fixed interaction order $k$. This observation motivates our goal of testing the presence of $k$-body interactions with QFI and product states. One should also mention that for $k=N$ the QFI in a product state is maximal, as in~\cite{Roy_2008}, and a constant trend for $N<2k$ is observed (see Fig.~\ref{fig:max_Fisher_fixed}).

A natural extension of the above considerations is to fix the number of qubits $N$ and study bounds on maximal QFI with Hamiltonians $H_k$ containing at most $k$-body couplings. Technically, knowing the eigenvalues of such Hamiltonians, see~(\ref{eq:eigenvalues}), we can once again maximise the variance and hence QFI directly. Here however we do not want to examine a sum of $k$-local Hamiltonians but its behaviour when the many-body couplings are varied. To detect interactions of order $k>2$ it is always sufficient to violate the bound for $\mathcal{H}^{(2)}=H_{1}+H_{2}$. Again, by solving~(\ref{eq:optimisation}) we can calculate the desired limits on QFI. As previously, the optimal solution has been found to be realised by $p_i=p$ for all $i$. 
For our problem, the exact calculations and numerical optimisation for up to $N=13$ give rise to the following extrapolated pattern 
\begin{equation*}
    \mathcal{B}_{1+2}=\max_{|\psi_{prod}\rangle}F[|\psi_{prod} \rangle ,\mathcal{H}^{(2)}] 
\end{equation*}
\begin{equation*}
=\max \Big( -\frac{16 N p (p-1) }{(N+1)^2}\lbrace (N - 2)^2 
\end{equation*}
\begin{equation}
+ 2(N - 1)p  \left[(2 N - 3) p - (2 N - 5)\right] \rbrace \Big),
    \label{eq:B12}
\end{equation}
and allow us to formulate the following criterion 
\begin{equation}
  \text{if }  F[\rho, \mathcal{H}^{(K)}]>\mathcal{B}_{1+2}\Rightarrow K \geq 3.
\end{equation}
The polynomial to be maximised is of the fourth order and a closed formula for the maxima can be found explicitly. However, due to its extensive structure, we chose to present the result in the above form. Explicit values of $\mathcal{B}_{1+2}$ for chosen $N$ are shown in Table.~\ref{table1}. It is worth noting that $\mathcal{B}_{1+2}$ does never exceed the maximal QFI for two-body interactions only and if needed, a stricter bound can be chosen.
Furthermore, the presented approach, in principle, can be performed for any $k$. Nevertheless, we will focus on the first physically interesting scenario presented above. 

\begin{table}
    \centering
    \begin{tabular}{|c|c|c|c|c|c|c|c|c|c|}
       \hline $N$ &  2 & 3 & 4& 5 & 6 & 7 & 8 & 9 & 10  \\
       \hline $\mathcal{B}_{1+2} $ & 1.78 & 2.68 & 3.61 & 4.57 & 5.53 & 6.51 & 7.49 & 8.47 & 9.46 \\
       \hline
    \end{tabular}
    \caption{Explicit values of the maximal quantum Fisher information $\mathcal{B}_{1+2}$~(\ref{eq:B12}) attainable in product states for a two-body Hamiltonian from the studied family. For each $N$, violation of this bound with any separable state yields the presence of at least three-body terms in the Hamiltonian.}
    \label{table1}
\end{table}

\subsubsection{Example}
\label{sec:example}

\begin{figure}
    \centering
    \includegraphics[width=0.48\textwidth]{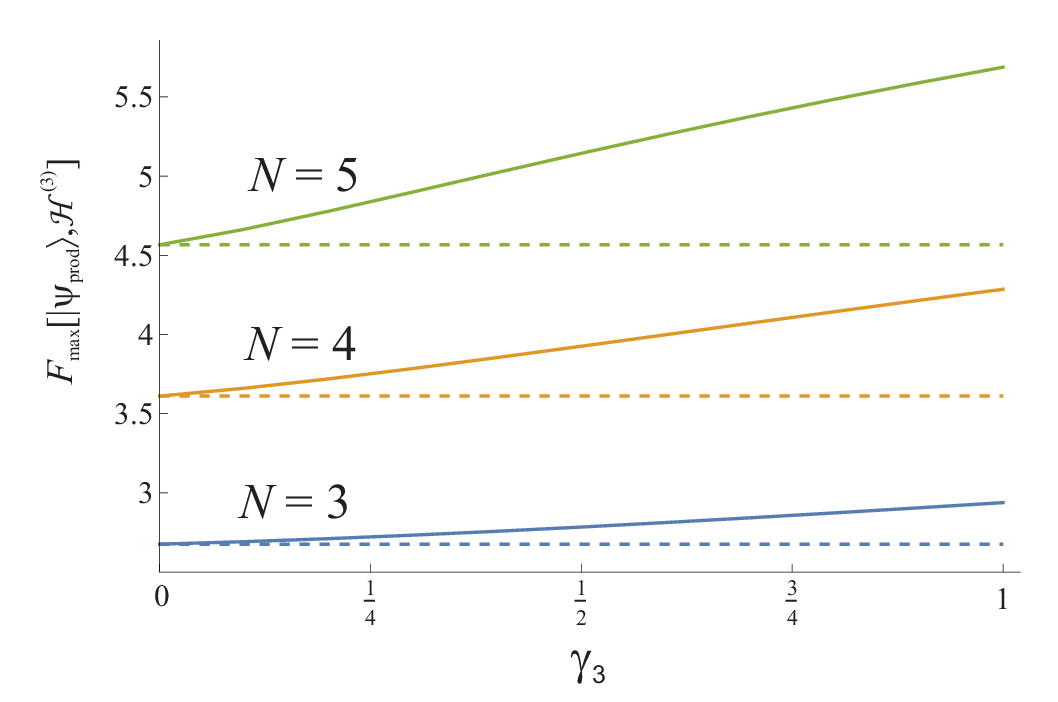}
    \caption{Maximal quantum Fisher information in a product state for Hamiltonian $\mathcal{H}_I$ ~(\ref{eq:ising}) with three-body contribution $\gamma_3 H_{3}$ for $N=3,4,5$ and varying $\gamma_3$ (solid line). Bounds $\mathcal{B}_{1+2}$, that allow one to verify the presence of higher-order interactions, were plotted with dashed lines. 
    One can see that it is possible to make statements about the order of interactions based on the presented results. If the bound of $\mathcal{B}_{1+2}$ is violated, then within the studied family of Hamiltonians, the interaction is at least $3$-local. 
}
    \label{fig:Fisher123}
\end{figure}

As an example, consider an Ising chain on a complete interaction graph with a uniform external field in the $z$ direction ($k\leq 2$). Tuning the field according to the coupling strength, as well as normalising an entire Hamiltonian we get
\begin{equation}
    \mathcal{H}_I= \mathcal{N} J \left( \sum_{i<j}^N \sigma^i_z \sigma^j_z + \sum_i^N \sigma_z^i \right),
    \label{eq:ising}
\end{equation}
where $\mathcal{N}=1/J(N+1)$. Suppose now that this system contains some amount of three-body interactions of the same symmetry, and the actual Hamiltonian is up to normalisation $\mathcal{H}^{(3)}=\mathcal{H}_I+\gamma_3 H_{3}$. In Fig.~\ref{fig:Fisher123} we plot the maximal QFI in product states for different $N$ and changing $\gamma_3$. Clearly, even for small $\gamma_3$, higher-order interaction is detected. Moreover, let us examine a case where we set different values of $\gamma_3= \lbrace 0.1, 0.5, 0.7, 1\rbrace$, and for each choice compute the QFI in a random pure three-qubit product state. In this scenario, it is also possible to violate $\mathcal{B}_{1+2}$. Namely, for $10^5$ samples the estimated frequency of violation is $\lbrace 0.23\%,4.3\%, 6.1\%,  8\% \rbrace$, for the respective choices of $\gamma_3$. Although, as expected, this frequency is small it is still significant, and one can make conclusions about the present interaction type. 

An interesting thing to comment on is the change in eigenlevels structure. In general, the state that maximizes variance is given as $1/\sqrt{2}(|E_{min}\rangle + |E_{max }\rangle)$. This however does not need to be a product state and, indeed, it is not in most of the cases. Nevertheless, if the order of interaction increases the structure of eigenlevels changes and Heisenberg scaling is available with product states. In fact, increasing $\gamma_3$ causes an attraction of the lowest energy levels resulting in stronger degeneracy when all couplings are equal. The resulting system is effectively a two-level structure and it is possible to form many product states which are a uniform superposition of two levels, hence the maximal value of variance and QFI can be achieved.

The discussed protocol could be especially useful for verifying if the many-body couplings have been engineered in a quantum simulator or another set-up without direct comparison of evolution with a specific $k$-local Hamiltonian. The importance of such tasks has been discussed in the introduction.

\section{Possible extensions}
\label{sec:extensions}

Here, we will shortly discuss the problem of $k$-body interactions verification outside of the discussed class of Hamiltonians. Let us consider a long-range interaction transverse field XY model
\begin{equation*}
    \mathcal{H}_{XY}=\mathcal{N}\lbrace J \sum_{i<j}\left[(1+\delta)\sigma_x^i \sigma_x^j+(1-\delta)\sigma_y^i \sigma_y^j \right]+b \sum_i \sigma_z^i \rbrace,
\end{equation*}
where $\mathcal{N}$ is a normalisation constant, $J$ exchange constant, $\delta$ anisotropy parameter and $b$ stands for an external magnetic field. For a similar discussion on the transverse field Ising model on a complete interaction graph see~\cite{Li_2019}. The above Hamiltonian differs significantly from the ones studied before and in principle, the conclusions drawn in the previous sections could not hold.

\begin{figure}
    \centering
    \includegraphics[width=0.4\textwidth]{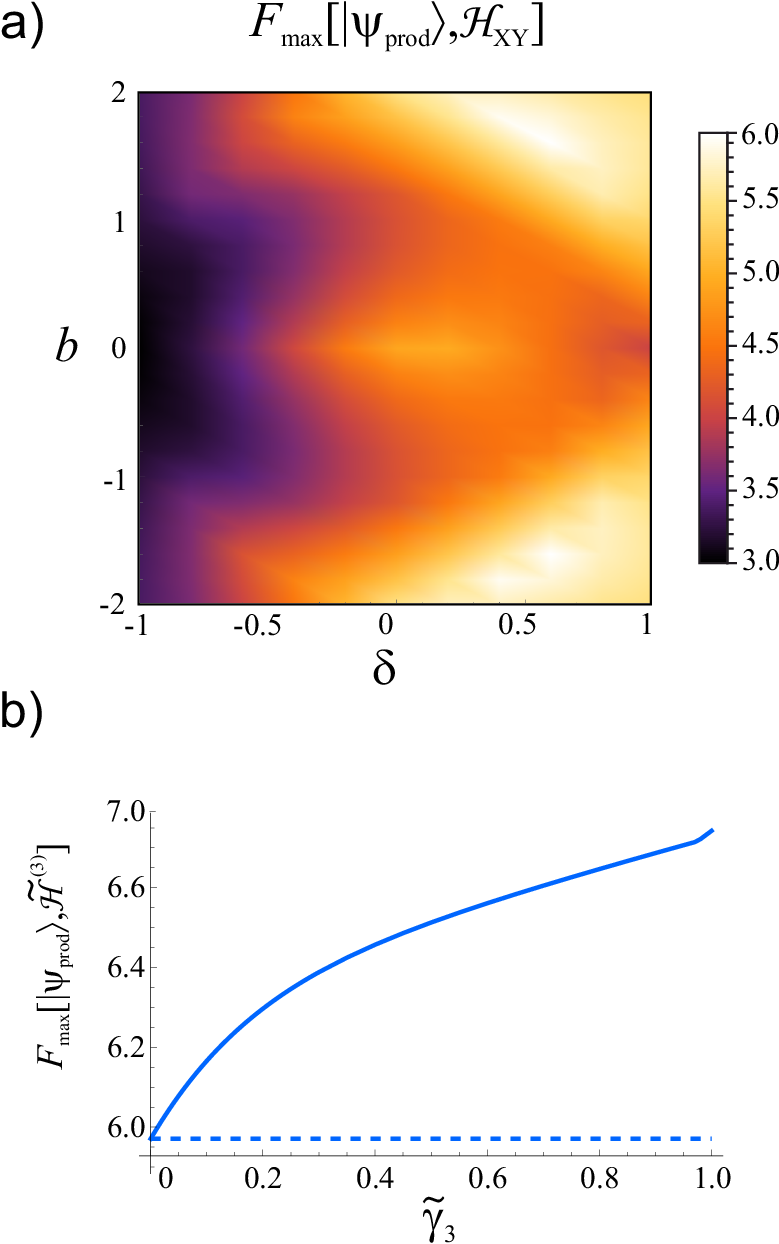}
     \caption{$a)$ Maximal quantum Fisher information in a three-qubit product state for the normalised transverse field XY Hamiltonian and different values of the external field $b$ and anisotropy parameter $\delta$. The free parameters were scanned within the region of $\pm 1$ for $\delta$ and $\pm 2$ for $b$ with a step of $1/5$. The maximal QFI in a product state was found to be $\mathcal{B}_{XY}=5.97$. For $\delta=0$, the XX model, it yields the maximal value of $5.51$. Since the QFI values plotted here do not saturate the upperbound of $N^2=9$, our reasoning can be used for the higher-order interactions verification.
    $b)$ Maximal quantum Fisher information in a product state for the normalised Hamiltonian $\Tilde{\mathcal{H}}^{(3)}=\mathcal{H}_{XY}+\Tilde{\gamma}_3\Tilde{H}_3$ with a varied $\Tilde{\gamma}_3$ and the optimal XY model parameters $\delta=0.6,b=-1.6$ (solid line). The value of $\mathcal{B}_{XY}$ is plotted with a dashed line. Its violation allows one to verify that a three-body interaction term is present in a system where one- and two-body terms are described with the XY model. Here, this is clearly possible. }
    \label{fig:XY}
\end{figure}

While working with the XY model we need to specify the free parameters. We choose to set $J=1$ and scan over different $\delta$ and $b$. The choice of $J$ is arbitrary due to the normalisation and scanning over different field and anisotropy values. The physical range of $\delta$ is $[-1,1]$. For large values of the external transverse field, the interaction terms contribution decreases and the results should converge to the case of local Hamiltonians. Thus, we chose to restrict $b$ to a significant region of $\pm \sqrt{J^2+\delta_{max}^2}$, i.e. $b \in [-2,2]$.  Performing a numerical optimisation over the set of pure product states for the case of $N=3$ we obtained the data plotted in Fig.~\ref{fig:XY}~a, where the step for the parameter change was chosen as $1/5$. Note that for $N=3$ particles, the long-range interaction term is equivalent to the periodic boundary condition and makes it more feasible experimentally. In general, we observe that the maximal QFI in a product state, i.e. $\mathcal{B}_{XY}$, is given as $5.97$. This corresponds to $\delta=0.6$ and $b=\pm 1.6$. For the $\delta=0$ cut (the XX model) the maximal QFI in a product state was found to be $5.51$. Both of these numbers are essentially smaller than $N^2=9$ which is needed to verify the higher-order interactions. Further in this section we will consider $\mathcal{H}_{XY}$ with the optimal parameters $\delta=0.6$ and $b=- 1.6$.

First, we will examine a specific example of the many-body interaction verification problem within the studied model. Then, a more general approach will be presented. Consider a three-body Hamiltonian $ \Tilde{H}_3$ which yields a QFI in a product state for the normalised $\Tilde{\mathcal{H}}^{(3)}=\mathcal{H}_{XY}+\Tilde{H}_3$ that violates the $\mathcal{B}_{XY}=5.97$. One possible choice is simply $\Tilde{H}_3 =  \sigma^1_x \sigma^2_x \sigma^3_x$, as it leads to the maximal QFI in a product state for $\Tilde{\mathcal{H}}^{(3)}$ equal to $6.74$. 
Similarly to the discussion in the previous section, we varied the amount of $\Tilde{H}_{3}$ contribution as a function of the coupling strength $\Tilde{\gamma}_{3}$. Once again, we conclude that the presence of higher-order interaction terms can be verified through QFI, as illustrated in Fig.~\ref{fig:XY}~b.

Choosing the three-body Hamiltonian in the XY model was far more arbitrary. Hence an additional interesting question arises. What if the underlying many-body interaction Hamiltonian could contain any three-body terms? For this reason, let us consider a three-body Hamiltonian of the general form $\Tilde{H}_3=\sum_{i,j,k=1}^3 \gamma_{ijk} \sigma_i \otimes \sigma_j \otimes \sigma_k$. Now, we introduce small random couplings by sampling $\gamma_{ijk}$ from a uniform distribution on $(-\gamma_{max},\gamma_{max})$ and choose $\gamma_{max}=0.5$. This leads to multiple violations of $\mathcal{B}_{XY}$ for $\mathcal{N}(\mathcal{H}_{XY}+\Tilde{H}_3)$ in the fixed optimal XY model product state with a frequency of $1.4\%$, estimated on $10^5$ runs. Consequently, it shows the ability of a more general three-body interaction verification within the XY model using the QFI.

\section{Conclusions}
We examined the possibility of detecting higher-order interactions with the use of quantum Fisher information. For normalised symmetric Ising-like Hamiltonians, we have shown that the maximal QFI in product states is ordered with respect to the fixed interaction order. Moreover, we calculated the maximal QFI obtained in the product states for the most natural scenario where one and two-body terms are present. This allowed us to verify the presence of at least three-body interactions in the chosen family of Hamiltonians through the violation of this bound. 
As a possible extension, we analysed an example concerning the three-body interactions verification in the XY model.

The considered problem has a strong foundational meaning as it paves the way for a better understanding of the nature of interactions. Furthermore, it encompasses some practical applications for Hamiltonian learning and could provide new perspectives for many-body interactions engineering. As argued before, we emphasise that QFI can be measured experimentally via multiple techniques. 

Future research on this problem could focus on generalising this observation to an arbitrary Hamiltonian class. We note that a different approach that contains no assumptions on the symmetry and interaction strength is possible. 
However, it would require the possibility of the mean energy measurement in an arbitrary state and a presumably unknown Hamiltonian. Very recently we became aware that a similar problem is being examined independently in a different manner by Bluhm \textit{et. al.} (see \cite{Bluhm_2024} for their preprint proposing a solution to this task).

\section{Acknowledgements} 
This research was supported by the National Science Centre (NCN, Poland) within the Preludium Bis project No. 2021/43/O/ST2/02679 (PC and WL) and the OPUS project No. 2023/49/B/ST2/03744 (TS).  For the purpose of Open Access, the authors have applied a CC-BY public copyright licence to any Author Accepted Manuscript version arising from this submission.

\appendix

\section{Maximal QFI in product states}
\label{app1}
To perform an exact maximisation of QFI for $N=3$ and $k=2$ let us consider the extrema conditions ~(\ref{eq:optimisation}) explicitly
\begin{equation*}
    \begin{cases}
        G_{2,3} [1 + 2 p_2 (p_3-1) - 2 p_3 + 2 p_1 (p_2 + p_3-1)]=0 \\
        G_{1,3} [1 + 2 p_2 (p_3-1) - 2 p_3 + 2 p_1 (p_2 + p_3-1)]=0 \\
        G_{1,2} [1 + 2 p_2 (p_3-1) - 2 p_3 + 2 p_1 (p_2 + p_3-1)]=0,
    \end{cases}
\end{equation*}
where $G_{i,j}=4 (1- p_i - p_j)$.
Discarding the minima solutions generated by eigenstates and limiting ourselves to $p_i \in [0,1]$, from $G_{i,j}$
we get
\begin{equation}
    p_1=p_2=p_3=\frac{1}{2},
\end{equation}
which is a local minima with QFI equal to 3. Assuming that the first term is non zero we obtain
\begin{equation}
    p_1=\frac{1 + 2 p_2 (1 - p_3) + 2 p_3}{2 (1 - p_2 - p_3)}.
    \label{eq_app:solution}
\end{equation}
This leads to QFI of 4 and can be also satisfied if all $p_i$ were taken to be equal. Indeed, taking $p_i=p$ for all $i$ we get
\begin{equation}
    F[|\psi \rangle ^{\otimes 3},H_{2}]= 2 \left( -18 p^4 +36 p^3 -24 p^2+3 p \right),
\end{equation}
with a maxima of 4 for $p=1/6(3+\sqrt{3})$.
For the three-body interaction Hamiltonian $H_{3}$ the calculations can be performed in an alternative manner. The spectrum of $H_{3}$ consists of two levels only. The energy of $-1/3$ is associated with the eigenstates $|111\rangle, |100\rangle, |010\rangle, |001\rangle$ and $|000\rangle, |110\rangle, |101\rangle, |011\rangle$ for the corresponding eigenvalue of $1/3$. The maximal algebraically allowed variance is given as the square of the energy bandwidth $(E_{max}-E_{min})^2$. Most often it is associated with highly entangled states, for the studied family of Hamiltonians a GHZ state. Here, however, due to additional degeneracies arising from the three-body couplings, see the discussion at the end of Sec.~\ref{sec:example}, a product state that saturates the variance can be constructed. Taking a uniform superposition of $|000\rangle$ and $|001\rangle$ we obtain $|0 \rangle \otimes |0\rangle \otimes 1/\sqrt{2}\,(|0\rangle + |1 \rangle)$. From the symmetry of the Hamiltonian any permutation of $1/\sqrt{2}\,(|0\rangle + |1 \rangle)$ among $|0\rangle $ is an equally valid solution.

Now, we give another example for $N=4$ and $k=2$. Here, we want to solve the set of four equations arising from~(\ref{eq:optimisation}). We do not report their explicit forms here, but one can easily generate them by calculating the variance in a parameterised product state. We found the following families of solutions to the given problem
\begin{subequations}
\begin{align}
   p_{i_1} = \frac{1}2 - \frac{1}{\sqrt{2}},\, p_{i_2} = \frac{1}2 - \frac{1}{\sqrt{2}},\, p_{i_3} = \frac{1}{2},\, p_{i_4} = \frac{1}{2} \\
    p_{i_1} = \frac{1}2 + \frac{1}{\sqrt{2}} , p_{i_2} = \frac{1}2 + \frac{1}{\sqrt{2}}, p_{i_3} = \frac{1}{2}, p_{i_4} = \frac{1}{2} \\
    p_{i_1} =  p_{i_2} = \frac{1}2 - \frac{1}{\sqrt{2}}, p_{i_3} =  p_{i_4} = \frac{1}2 + \frac{1}{\sqrt{2}}\\
    p_1=p_2=p_3=p_4=\frac{1}{2} \\
    p_{i_1} = \frac{1}2 - \frac{1}{\sqrt{10}} , p_{i_2} = \frac{1}2 - \sqrt{\frac{5}{2}}, p_{i_3} = \frac{1}{2}, p_{i_4} = \frac{1}{2} \\
    p_{i_1} = \frac{1}2 - \frac{1}{\sqrt{10}} , p_{i_2} = \frac{1}2 - \sqrt{\frac{5}{2}}, p_{i_3} = \frac{1}{2}, p_{i_4} = \frac{1}{2} \\
    p_{i_1} = \frac{1}2 + \frac{1}{\sqrt{10}} , p_{i_2} = \frac{1}2 + \sqrt{\frac{5}{2}}, p_{i_3} = \frac{1}{2}, p_{i_4} = \frac{1}{2} \\
     p_1 =p_2=p_3=p_4= \frac{1}2 \pm \frac{1}{\sqrt{10}}
\end{align}
\end{subequations}
where the trivial (eigenstate) solutions were discarded. These families are indexed with $(i_1,i_2,i_3,i_4)$, which take distinct values from $[1,4]$. This resembles the symmetric structure of the studied Hamiltonians. One can check directly that the last solution with $\forall_i p_i=p$ gives rise to the highest QFI as reported in the main text, i.e., $F_{max}[|\psi_{prod} \rangle ,H_{2}]=24/5=4.8.$

For $N>3$ solutions of the form ~(\ref{eq_app:solution}) do not guarantee the vanishing derivatives for all parameters, as shown for $N=4$. Nevertheless, the more complicated solutions also contain the one for which the optimal state is a tensor product of single-qubit states (see the main text). To further check our results and examine $N \geq 5$, we found the upper bound on the QFI with standard numerical and symbolical optimisation techniques built in Wolfram Mathematica 13.3. All of the results obtained by solving the extrema conditions were in agreement with the numerical calculations. It is worth noting that the problem under consideration is exactly solvable. One can express the variance of a Hamiltonian on two state copies, i.e., $\Tr[(\openone \otimes H^2-H \otimes H)\rho']$ with $\rho'=\rho \otimes \rho$, which reduces it to a linear form. Constraints on the reduced states can ensure that the maximisation is performed over the separable states $\rho$ and hence solving the problem.

\bibliographystyle{apsrev4-2}
\bibliography{ref_hamiltoniany.bib}

\end{document}